\documentclass[sigconf]{acmart}
\AtBeginDocument{%
  }

\setcopyright{acmlicensed}
\copyrightyear{2026}
\acmYear{2026}
\acmDOI{XXXXXXX.XXXXXXX}
\acmConference[L@S '26]{the 13th ACM Conference on Learning @ Scale}{June 29 -- July 3, 2026}{Seoul, South Korea}
\acmISBN{978-1-4503-XXXX-X/2026/06}

\begin{document}


\title{PromptDecipher: Supporting AI Tutor Authoring Through Editable Simulated Interactions}

\author{Miina Koyama}
\affiliation{%
  \institution{Carnegie Mellon University}
  \city{Pittsburgh}
  \state{Pennsylvania}
  \country{USA}}
\email{mkoyama@cs.cmu.edu}

\author{Ruiwei Xiao}
\affiliation{%
  \institution{Carnegie Mellon University}
  \city{Pittsburgh}
  \state{Pennsylvania}
  \country{USA}}
\email{ruiweix@cs.cmu.edu}

\author{John Stamper}
\affiliation{%
  \institution{Carnegie Mellon University}
  \city{Pittsburgh}
  \state{Pennsylvania}
  \country{USA}}
\email{jstamper@cs.cmu.edu}

\renewcommand{\shortauthors}{Miina Koyama, Ruiwei Xiao, and John Stamper}

\begin{abstract}
Chatbots have long been explored as tools to support learning ~\cite{Hwang23}, and recent advances in large language models have significantly expanded the availability of platforms for educators to author AI tutoring chatbots ~\cite{Kasneci23, Yoo25}. Yet effective authorship demands more than writing a system prompt; it requires educators to act as learning designers, AI interaction designers, and QA engineers. In practice, however, teachers rarely fulfill these roles. Our formative study found that virtually none systematically tested their bots before deploying them to students. To address this gap, we present PromptDecipher, a system that restructures the authoring workflow around a direct correction-based interaction rather than writing abstract system prompts, teachers interact with a live chat preview and edit undesirable bot responses. An automated pipeline then analyzes the correction, proposes a targeted system prompt rewrite, and validates the change across pre-defined test scenarios. This enforces QA as a first-class activity and scaffolds teachers in roles they would otherwise skip. PromptDecipher will be deployed in an \textit{AI for Educators} course enrolling hundreds of higher-education instructors. A live prototype\footnote{\url{https://teacher-prompting.vercel.app/}}, an anonymized codebase\footnote{\url{https://anonymous.4open.science/r/teacher-prompting-2EDF/}}, and anonymized demo\footnote{\url{https://tinyurl.com/las-prompt-decipher-demo}} are available via links in the footnote.
\end{abstract}

\ccsdesc[500]{Applied computing~Interactive learning environments}
\ccsdesc[300]{Human-centered computing~Interactive systems and tools}

\keywords{AI in Education, Generative AI, Authoring Tools, Prompt Engineering, Educational Technology}

\begin{teaserfigure}
  \centering
  \includegraphics[width=0.86\textwidth]{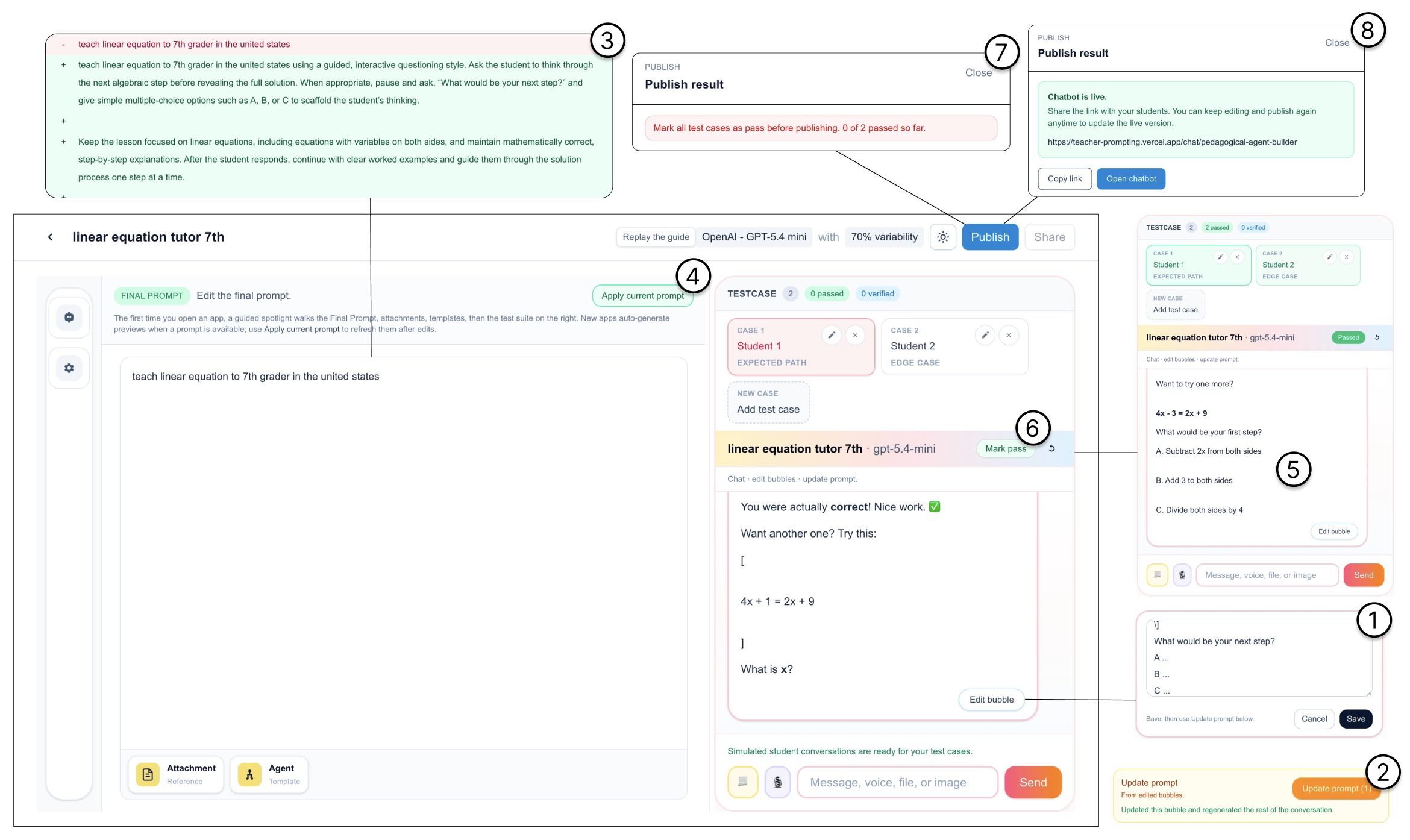}
  \caption{The PromptDecipher workflow. (1) A teacher edits an unsatisfactory bot response in the test case chat. (2) The system surfaces a prompt update button derived from the correction. (3) The AI automatically revises the system prompt, visualized as a tracked diff. (4) The teacher applies the updated prompt to the editor. (5) The bot's response in the test case is updated, reflecting the prompt change. (6) If the response looks good, the teacher clicks 'Mark pass' to approve the test case. (7) Publication is blocked until all test cases are marked as passed. (8) Once all cases pass, the bot goes live and a shareable link is generated.}
  \Description{A screenshot of the PromptDecipher system with three annotated regions. The main panel shows the prompt editor on the left displaying a tracked diff with red deleted lines and green added lines, and the test case chat panel on the right where a teacher can edit bot responses. An annotation reads 'Editing a bot response automatically generates a diff and proposes a prompt update.' To the right, two publish dialog boxes are shown: the top one displays an error message blocking publication because not all test cases have passed, annotated as 'Publication is blocked until all test cases pass.' The bottom one shows a success state with a live chatbot link, annotated as 'Once all cases pass, the bot is published and a live link is generated.'}
  \label{fig:teaser}
\end{teaserfigure}


\maketitle

\section{Introduction}
In this paper, we present PromptDecipher, a web-based system that enables educators to author and refine AI tutoring chatbots through direct correction of bot responses, without requiring expertise in prompt engineering. Chatbot authoring platforms such as Playlab\footnote{\url{https://www.playlab.ai/}} allow teachers to create AI tutors via system prompt editing, but creating an educationally effective bot demands that teachers simultaneously act as learning designers, AI interaction designers, and QA engineers, which are roles far beyond their typical experience ~\cite{xie25}. Prior work on end-user prompt engineering has documented that non-experts approach prompting opportunistically rather than systematically, and struggle to translate observations about undesired outputs into concrete prompt requirements ~\cite{JD23, Ma25}. Our formative study of 121 chatbots created by instructors in an "AI for Educators" MOOC corroborates this at scale: while teachers successfully specified learning content, nearly all failed to engage in any systematic testing before publication — a critical concern when bots are deployed to real learners, including K–12 students~\cite{Yoo25}.
The core problem is a mismatch between the interface these platforms provide — a raw text editor for a system prompt — and teachers' existing mental models. Teachers are deeply experienced in giving corrective feedback on student work, but have no prior frame for specifying AI behavior in natural language. PromptDecipher addresses this by repositioning the core authoring activity: instead of writing a prompt, teachers correct chatbot responses. Submitting a correction triggers an automated pipeline that (1) identifies what changed and why, (2) proposes a minimal rewrite of the underlying system prompt, and (3) validates the updated prompt against a suite of test scenarios. Because teachers must complete at least one such cycle before publishing, QA is structurally embedded in the workflow.

\section{System Overview}
In PromptDecipher, a teacher creates a new bot, selects a foundation model (OpenAI, Anthropic, or Google), and optionally uploads course materials. Rather than writing a system prompt from scratch, the teacher is directed to the test environment: a simulated student chat. They select a student profile (e.g., "expected path," "struggling learner," "off-topic input"), read the bot's response, and either mark it as passing or edit it to reflect the desired behavior. Upon submitting an edit, the Reverse Prompting Pipeline is triggered:

\begin{enumerate}
    \item \textbf{Diff analysis}. An LLM compares the original and corrected responses to infer the teacher's pedagogical intent (e.g., "the bot should ask a follow-up question rather than give the answer directly").
    \item \textbf{Prompt rewrite}. A targeted addition or modification to the system prompt is proposed and shown to the teacher for review.
    \item \textbf{Regression verification}. The revised prompt is automatically evaluated across all previously passed test cases; any regression is flagged for the teacher's attention before they can proceed.
\end{enumerate}

This test-correct-verify cycle continues until the teacher is satisfied; publication is gated behind at least one completed cycle. The system additionally supports direct prompt editing via templates and an AI-assisted discussion panel, though the correction-based pipeline is the primary contribution of this demonstration.

\section{Demonstration and Interaction Plan}
Attendees will author their own AI tutoring bot end-to-end using provided laptops:

\begin{enumerate}
    \item \textbf{Setup (2 min)}. The attendee creates a new bot, enters a brief description of the learning context (e.g., "a Socratic tutor for introductory statistics"), and selects a foundation model.
    \item \textbf{Test and correct (5 min)}. The attendee selects a simulated student profile and reads the bot's initial response. They then edit the response to reflect what they wish the bot had said and submit the correction. The Reverse Prompting Pipeline runs live, showing the inferred intent, the proposed prompt update, and the regression check results.
    \item \textbf{Iterate and publish (3 min)}. The attendee runs one additional scenario, makes any further corrections, and publishes the bot. A shareable link is generated; attendees can immediately interact with their published bot as a student.
\end{enumerate}

Demo team members will be present throughout to guide attendees and collect informal feedback. A poster adjacent to the station will provide context on the formative study and design rationale.

\section{Discussion and Conclusion}
PromptDecipher demonstrates how authoring interface design can shape educator behavior at scale. By making the modification directly on the simulated tutoring chat — not the prompt — the primary unit of authoring, the system transforms an abstract engineering task into a familiar pedagogical one, and embeds quality assurance structurally rather than relying on voluntary compliance. PromptDecipher is scheduled for deployment in an "AI for Educators" MOOC with hundreds of higher-education instructors in fall 2026, where collected usage data will allow us to examine whether correction-based authoring increases testing rates and improves prompt quality relative to conventional workflows. Future directions may include auto-generation of additional edge-case scenarios and integration of learning science guidance to help teachers recognize pedagogical strategies surfaced through their own corrections.

\bibliographystyle{ACM-Reference-Format}
\bibliography{references}
\end{document}